\documentclass[a4paper,reqno]{amsart}
\usepackage{amssymb}
\usepackage{latexsym}
\usepackage{amsmath}
\usepackage{euscript}
\usepackage{graphics,color}
\usepackage[all]{xy}
\usepackage[margin=3cm]{geometry}
\usepackage{MnSymbol} % for \udots
\usepackage{tikz}
\usepackage[normalem]{ulem}

\def\cJ{{\mathcal J}}

\newcommand{\be}{\begin{equation}}
\newcommand{\ee}{\end{equation}}
\newcommand{\ba}{\begin{eqnarray}}
\newcommand{\ea}{\end{eqnarray}}
\newcommand{\baa}{\begin{eqnarray*}}
\newcommand{\eaa}{\end{eqnarray*}}
\newcommand{\bb}{\begin{thebibliography}}
\newcommand{\eb}{\end{thebibliography}}

\newcounter{my}
\newcommand{\he}%
   {\stepcounter{equation}\setcounter{my}%
   {\value{equation}}\setcounter{equation}0%
   }%
\newcommand{\she}%
   {\setcounter{equation}{\value{my}}%
    }%

\theoremstyle{definition}

\numberwithin{equation}{section}

\newcommand{\hg}[2]{\,\mbox{}_{#1}F_{ #2}\!}

\newcommand{\argu}[3]{\left(\begin{array}{c} #1\\#2\end{array} ; #3\right)}

\DeclareRobustCommand{\erase}{\bgroup\markoverwith{\textcolor{red}{\rule[.5ex]{2pt}{0.8pt}}}\ULon}

\title{The rank two Jacobi algebra}
\author{Nicolas Crampé, Satoshi Tsujimoto, Luc Vinet, Alexei Zhedanov}
\address{CNRS - Université de Montréal CRM-CNRS, P.O. Box 6128, Centre-ville Station, Montr\'eal (Qu\'ebec), H3C 3J7}

\address{Graduate School of Informatics, Kyoto University,
Yoshida-Honmachi, Kyoto, Japan 606-8501}

\address{IVADO and Centre de recherches math\'ematiques, Universit\'e de Montr\'eal, P.O. Box 6128, Centre-ville Station, Montr\'eal (Qu\'ebec), H3C 3J7}

\address{Department of Mathematics, School of Information, Renmin University of China, Beijing 100872,CHINA}

\begin{document}

\begin{abstract}
The quadratic rank two Jacobi algebra is identified from the relations obeyed by the bispectral operators of the two variable Jacobi polynomials orthogonal on the triangle. It is seen to admit as subalgebras Racah and Jacobi algebras of rank one. The dual realizations in terms of differential operators in the variable representation and in terms of difference operators in the degree representation are provided. Structure relations for the two variable Jacobi polynomials are obtained as a by product.
\end{abstract}

\maketitle

\section{Introduction}

The rank two Jacobi algebra will be defined from the relations obeyed by the operators involved in the bispectral relations of the two variable Jacobi polynomials orthogonal on the triangle. Its dual representations in both variables and degrees will naturally provide structure relations for the bivariate polynomials.

Orthogonal polynomials and special functions with rich sets of properties typically entail algebraic structures that provide new tools to describe symmetries and these algebras often find applications in many different areas. This happens in particular when considering the families of univariate orthogonal polynomials of the Askey scheme. These polynomials are bispectral as they obey two eigenvalue equations: the three-term recurrence relation required by their orthogonality where they appear as eigenfunctions of a difference operator acting on the degree index with an eigenvalue that is a function of the variable and another giving them as eigenfunctions of a differential or difference operator acting on the variable with the eigenvalue depending in this case on the degree. The algebras to be discussed are realized by these so-called bispectral operators (either in the variable or the degree representation) and they hence encode the bispectrality of the underlying polynomials. The simplest example beyond Lie algebras is offered by the Jacobi polynomials \cite{koekoek2010hypergeometric}. Indeed, denoting by $K_1$ the hypergeometric operator of which the Jacobi polynomials with parameters $\alpha$ and $\beta$ are eigenfunctions and by $K_2$ the multiplication by the variable, it has been observed (see for instance \cite{genest2016tridiagonalization} and Appendix B) that the following relations are realized:
\begin{align} \label{rank1}
  [K_1,[K_1,K_2]]=&\;-2\{K_1,K_2\}+2 K_1+(\alpha+\beta)(\alpha+\beta+2)K_2-(\alpha+\beta)(\alpha+1),\nonumber\\
  [K_2,[K_2,K_1]]=&\;-2K_2^2+2K_2,
\end{align}
with $[A,B]=AB-BA$ and $\{A,B\}=AB+BA$. This defines the rank one Jacobi algebra  with its characteristic quadratic feature. A similar approach can be adopted with the bispectral operators of the Wilson and Askey-Wilson polynomials (or equivalently with those of the Racah and $q$-Racah polynomials for what concerns the algebras) that sit at the top of the Askey tableau and its $q$-version \cite{koekoek2010hypergeometric}. This leads to the eponymous Racah \cite{zhedanov1988nature, genest2014superintegrability,genest2014racah} and Askey-Wilson \cite{zhedanov1991hidden, granovskii1993hidden, crampe2021askey} algebras respectively \footnote{It is customary to name algebras after the polynomials to which they are associated.}. The Jacobi algebra is readily found to be a special case of the Racah algebra by setting certain parameters in the latter to specific values. Beyond capturing the properties of the Jacobi polynomials through its representations \cite{granovskii1992mutual}, the Jacobi algebra has been shown to be the dynamical algebra of many exactly solvable quantum mechanical models \cite{granovskii1993hidden,lutsenko1992jacobi}. We here wish to introduce and describe in detail the rank two generalization of this quadratic algebra by following the path that initially led to the rank one structures, that is by anchoring the identification of the algebra in a two variable generalization of the Jacobi polynomials.

Before we introduce the framework for that, let us indicate that Racah and Askey-Wilson algebras of higher ranks have already been defined and connected to families of multivariate orthogonal polynomials. We shall here limit this part of the background to the Racah story, i.e. to the situation with $q=1$. Basically, the higher rank extensions have proceeded as follows. Key was the observation, using in particular the generic superintegrable model on the 2-sphere \cite{genest2014superintegrability}, that (central extensions) of the rank one Racah algebra can be realized by the intermediate Casimir operators in the tensor product of three $sl_2$ representations; furthermore, the univariate Racah polynomials were then seen to occur naturally as the overlaps between different eigenbases of these generators. Extending this construction to tensor products with more than three factors \cite{de2017higher,crampe2021racah}, paved the way towards the generalizations to ranks larger than one that were being looked for and provided in particular algebraic interpretations for the multivariate orthogonal polynomials of discrete variables introduced by Tratnik \cite{tratnik1991some}.
This was explored concretely for the Racah algebra of rank two in a superintegrable model on the 3-sphere \cite{kalnins2011two} and culminated with the construction of the representations of this algebra in \cite{crampe2023representations},\cite{post2024racah}. Like for rank one, it is expected that the rank two Jacobi algebra we wish to spell out will also be a specialization of this rank two Racah algebra.

Some 50 years ago, Koornwinder \cite{koornwinder1975two} designed an approach, somewhat similar to the Tratnik method in the discrete case, to construct orthogonal polynomials in two variables and he offered examples of such families that obeyed a pair of differential equations. Among them were the two variable Jacobi polynomials orthogonal on the triangle that had been introduced previously by Proriol \cite{proriol1957famille}, see also \cite{dunkl2014orthogonal}, \cite{karniadakis2005spectral}. They had also appeared in the classification performed by Krall and Sheffer \cite{krall1967orthogonal} as well as by Engelis \cite{engelis1974some} of the orthogonal polynomials of degree $n$ in two variables that are eigenfunctions of a second order partial differential operator. They were further shown in \cite{vinet2003two} to be the wave functions of a quantum superintegrable model on the two-dimensional (pseudo)sphere. This generated a lot of research as the review \cite{fernandez2010recent} testifies and the interest has not waned. In their study of the bispectrality properties of the (discrete) Tratnik polynomials, Geronimo and Iliev \cite{geronimo2010bispectrality} examined these aspects of the multivariate Jacobi polynomials through limits of the Hahn polynomials in many variables. Certain $q$-analogs of the two-variable Jacobi polynomials have also been constructed
: the bivariate little $q$-Jacobi ones in \cite{dunkl1980orthogonal} which were generalized to the bivariate big $q$ Jacobi polynomials in \cite{lewanowicz2010two}. Structure relations for the latter were obtained in \cite{lewanowicz2013structure} and their $q \rightarrow 1$ limits provided the corresponding relations for the ordinary two variable Jacobi polynomials. Furthermore, the $q \rightarrow -1$ limit of the bivariate big $q$-Jacobi polynomials was carried out in \cite{genest2015two} to identify the bivariate big $-1$ Jacobi polynomials \cite{vinet2012limit} and their bispectral features. The bivariate Jacobi polynomials also appear in \cite{iliev2018symmetry} as wavefunctions of the generic superintegrable quantum model on the 2-sphere and hence as basis vectors for representations of the rank one Racah algebra. See also \cite{labriet2024realisations} for realizations of the Racah algebra in terms of Jacobi operators and convolution relations involving two-variable Jacobi polynomials as well as \cite{de2017higher} and \cite{de2019bargmann} for other occurrences of these polynomials in connection with the Racah algebra.

The objective here is to propose a straightforward definition of the rank two  Jacobi algebra in terms of generators and relations from computing those satisfied by the operators involved in the bispectral equations of the two variable Jacobi polynomials. This will supplement significantly and clearly the corpus of knowledge specific to these polynomials and will stand to generate further development. The paper will unfold as follows. In Section 2, we recall the definition of the two variable Jacobi polynomials that are orthogonal on the triangle and which are the central characters of this study. The focus is put on their bispectrality properties that form the starting point of this investigation. The reader will find in Section 3 the relations that we shall take to define the rank two Jacobi algebra. The presentation will be obtained by pursuing the closure in quadratic expressions of a finite set of generators of the multiple commutators between the bispectral operators taken to be the two differential operators of which the bivariate Jacobi polynomials are eigenfunctions and, viewed as corresponding to the recurrence relations, the multiplication by two linear and independent combinations of the variables. The resulting rank two algebra will be seen to have as subalgebras the rank one Racah algebra as well as various rank one Jacobi algebras. The dual realization of the rank two Jacobi algebra in terms of difference operators acting on the degrees will be given in Section 4. In effect, the results of this section will amount to providing the representation of the rank two Jacobi algebra on the basis formed by the set of two variable Jacobi polynomials. Structure relations for these polynomials will thus be identified. In a final section, various possible follow-up projects will be evoked. A compendium of relevant properties of the univariate Jacobi polynomials will be found in Appendix A  with a return to the rank one Jacobi algebra relations \eqref{rank1}. Finally, many useful differential properties of the two variable Jacobi polynomials that are probably new  are collected in Appendix B.

\section{The two variable Jacobi polynomials on the triangle and their bispectral relations}
This section introduces the bivariate Jacobi polynomials that are orthogonal on the triangle and records their bispectral properties.
\subsection{Definition and orthogonality}
The two-variable Jacobi polynomials $J_{n,k}^{(a,b,c)}(x,y)$ orthogonal on the triangle \cite{proriol1957famille, koornwinder1975two} of total degree $n+k$ in the variables $x$ and $y$, are given as follows  
\begin{align}
 &J_{n,k}^{(a,b,c)}(x,y)=J_{n-k}^{(a,b+c+2k+1)}\bigl(x\bigr) \cdot (1-x)^k \cdot J_k^{(b,c)}\left(\frac{y}{1-x}\right) \quad \mbox{}\quad (n \ge k \ge 0),
\end{align}
in terms of the transformed univariate Jacobi polynomial $J_n^{(a,b)}(x)$ on the interval $[0,1]$ whose definition and some properties will be found in Appendix A.
The polynomials $J_{n,k}^{(a,b,c)}(x,y)$ satisfy the following orthogonality relation for $a,b,c > -1$:
\begin{align}
 \int_{0\le x\le  1-y \le 1} J_{n,k}^{(a,b,c)}(x,y) J_{n',k'}^{(a,b,c)}(x,y) x^a y^b (1-x-y)^c dx dy = h_{n,k} \delta_{n,n'} \delta_{k,k'} 
\end{align}
with
\begin{equation}
    h_{n,k}= \dfrac{(b+c+2k+2)_{n-k}(b+c+k+1)_k}{(a+b+c+2n+2) \, (n-k)! \, k!}\dfrac{\Gamma(a+n-k+1)\,\Gamma(b+k+1)\,\Gamma(c+k+1)}{\Gamma(a+b+c+n+k+2)}.
\end{equation}

\subsection{Differential equations}
The two variable Jacobi polynomials are solutions of two differential eigenvalue equations that we will now describe. Consider the operator $L^{(a,b,c)}$ given by
\begin{align} \label{L}
  &L^{(a,b,c)}=\nonumber\\
  &x(1-x)\partial_{xx}+y(1-y)\partial_{yy}-2xy\partial_{xy}+(a+1-(a+b+c+3)x)\partial_{x}+(b+1-(a+b+c+3)y)\partial_{y},
\end{align}
where $\partial_x f(x,y)=\frac{ \partial}{\partial x} f(x,y)$ etc.
This operator admits symmetries. Indeed, it can be written as the\ sum of three operators $L_i, i=1,2,3 $ that each commutes with $L^{(a,b,c)}$:
\begin{equation}
    L^{(a,b,c)} = L_1^{(b,c)}+L_2^{(a,c)}+L_3^{(a,b)}, \quad [L^{(a,b,c)},L_1^{(b,c)}]=[L^{(a,b,c)},L_2^{(a,c)}]=[L^{(a,b,c)},L_3^{(a,b)}]=0,
\end{equation}
with 
\begin{align}
&L_1^{(b,c)}=((b+1)(1-x)-(b+c+2)y)\partial_y + y(1-x-y) \partial_{yy},\label{Ly}\\
&L_2^{(a,c)}=((a+1)(1-y)-(a+c+2)x)\partial_x + x(1-x-y) \partial_{xx},\label{Lx}\\
&L_3^{(a,b)}=((a+1)y-(b+1)x)(\partial_x-\partial_y) + xy \left(\partial_{xx}+\partial_{yy}-2\partial_{xy}\right)\label{Lz}.
\end{align}
It is found that the two-variable Jacobi polynomials are joint eigenfunctions of 
$L^{(a,b,c)}$ and $L_1^{(b,c)}$ with eigenvalues given in the two equations below: 
\begin{align}
 &L^{(a,b,c)}[J_{n,k}^{(a,b,c)}(x,y)]=-n\,(n+a+b+c+2)J_{n,k}^{(a,b,c)}(x,y),\label{bipde1}\\
 &L_1^{(b,c)}[J_{n,k}^{(a,b,c)}(x,y)]=-k\,(k+b+c+1)J_{n,k}^{(a,b,c)}(x,y) \label{bipde2}.
\end{align}
In the following (except in the appendices) we shall suppress the superscripts with the letters to make the notation less cluttered.
\begin{proof} 
For derivations of \eqref{bipde1} and \eqref{bipde2}, the reader may consult the literature, e.g. \cite{koornwinder1975two} and references therein or \cite{geronimo2010bispectrality} where the bispectral properties of the multivariate Jacobi polynomials are obtained from those of the multivariate Hahn polynomials through a limit procedure. For the sake of completeness we offer here indications of how these equations can be obtained directly. This makes use of the differential formulas given in Appendix B.

By observing that $s_1^{*}-1$  shifts the parameters of the two variable Jacobi polynomials from $(a, b+1, c-1)$ to $(a, b, c)$, it is straightforward to show that the operator $L_1$, which is given by the product $(s_1^{*}-1) s_1 $ plus a constant, acts as a scalar multiplication on the two variable Jacobi polynomials. This gives a direct proof of (\ref{bipde2}). Similarly, since $s_2^{*}-1$ and $s_3^{*}-1$ shift the parameters from $(a-1, b, c+1) \to (a, b, c) $ and $ (a+1, b-1, c) \to (a, b, c) $, respectively, it follows that these operators  $(s_2^{*}-1) s_2$ and $(s_3^{*}-1) s_3$ map $J_{n,k}^{(a,b,c)}$ into the span  of 
\begin{align}
    \operatorname{span} \{ J_{n,k-1}^{(a,b,c)}, J_{n,k}^{(a,b,c)}, J_{n,k+1}^{(a,b,c)} \}.   
\end{align}
Taking the sum of these actions yields the equation $ (s_2^{*}s_2+s_3^{*}s_3 -s_2- s_3)J_{n,k}^{(a,b,c)}(x,y) = \lambda J_{n,k}^{(a,b,c)}(x,y)$ with some constant $\lambda$. Finally, by using \eqref{L:s1s2s3}, one can establish the differential equation (\ref{bipde1}).
\end{proof}
\subsection{Recurrence relations}\label{sec:RR1}
We now turn to the two recurrence relations of the two variable Jacobi polynomials $J_{n,k}^{(a,b,c)}(x,y)$ which respectively involve three and nine terms and read:
\begin{align}
x& J_{n,k}^{(a,b,c)}(x,y) =\Hat{X}_1 J_{n,k}^{(a,b,c)}(x,y),\label{RR1}
\end{align}
\begin{align}
(1-x-y) &J_{n,k}^{(a,b,c)}(x,y) = \Hat{X}_3\, J_{n,k}^{(a,b,c)}(x,y)
 \label{RR2}
\end{align}
where
\begin{align}
\Hat{X}_1&=-
\dfrac{(n-k+1)(n+k+a+b+c+2)}{(2n+a+b+c+2)(2n+a+b+c+3)} S_{+}\nonumber\\
&+\dfrac12\bigg(1-\dfrac{(2k+a+b+c+1)(2k-a+b+c+1)}{(2n+a+b+c+1)(2n+a+b+c+3)} \bigg)I\nonumber\\
&-\dfrac{(n-k+a)(n+k+b+c+1)}{(2n+a+b+c+1)(2n+a+b+c+2)}S_{-},\label{M2}
\end{align}
\begin{align}
    \Hat{X}_3 &=
\dfrac{1}{2} I - \dfrac{1}{2} \Hat{X}_1    \nonumber\\&
+\dfrac{(k+1)(k+b+c+1)(n+k+a+b+c+2)(n+k+a+b+c+3) }{(2k+b+c+1)(2k+b+c+2)(2n+a+b+c+2)(2n+a+b+c+3)} 
S_{+}T_{+}\nonumber\\
&+\dfrac{2(k+1)(k+b+c+1)(n-k+a)(n+k+a+b+c+2) }{(2k+b+c+1)(2k+b+c+2)(2n+a+b+c+1)(2n+a+b+c+3)} 
T_{+}\nonumber\\
&+\dfrac{(k+1)(k+b+c+1)(n-k+a-1)(n-k+a) }{(2k+b+c+1)(2k+b+c+2)(2n+a+b+c+1)(2n+a+b+c+2)} 
S_{-}T_{+}\nonumber\\
&+\dfrac{(c^2-b^2)(n-k+1)(n+k+a+b+c+2) }{2(2k+b+c)(2k+b+c+2)(2n+a+b+c+2)(2n+a+b+c+3)}
S_{+}\nonumber\\
%%%
&+\frac{c^2-b^2}{4}\bigg(\dfrac{1}{(2k+b+c)(2k+b+c+2)}+\dfrac{1}{(2n+a+b+c+1)(2n+a+b+c+3)}
\nonumber\\
& \qquad  + \dfrac{1-a^2 }{(2k+b+c)(2k+b+c+2)(2n+a+b+c+1)(2n+a+b+c+3)} \bigg)  
I \nonumber\\
%%%
&+\dfrac{(c^2-b^2)(n-k+a)(n+k+b+c+1) }{2(2k+b+c)(2k+b+c+2)(2n+a+b+c+1)(2n+a+b+c+2)} 
S_{-}\nonumber\\
&+\dfrac{(k+b)(k+c)(n-k+1)(n-k+2) }{(2k+b+c)(2k+b+c+1)(2n+a+b+c+2)(2n+a+b+c+3)} 
S_{+}T_{-}\nonumber\\
&+\dfrac{2(k+b)(k+c)(n-k+1)(n+k+b+c+1) }{(2k+b+c)(2k+b+c+1)(2n+a+b+c+1)(2n+a+b+c+3)} 
T_{-}\nonumber\\
&+\dfrac{(k+b)(k+c)(n+k+b+c)(n+k+b+c+1) }{(2k+b+c)(2k+b+c+1)(2n+a+b+c+1)(2n+a+b+c+2)} 
S_{-}T_{-} \label{M}
\end{align}
with the shift operators $S_{+},S_{-}, T_{+}, T_{-}$:
\begin{align}\
    &S_{+} f_{n,k} = f_{n+1,k}, \quad 
    S_{-} f_{n,k} = f_{n-1,k}, \quad 
    T_{+} f_{n,k} = f_{n,k+1},  \quad 
    T_{-} f_{n,k} = f_{n,k-1}. \label{ST}
\end{align}

\begin{proof}
The first relation \eqref{RR1} is readily obtained from the recurrence relation \eqref{RR:univariate} of the univariate Jacobi polynomials. The proof of
(\ref{RR2}) requires the use of the contiguity relations of the Jacobi polynomials $J_n^{(a,b)}(x)$ that are  given in Appendix A. The derivation goes as follows. We have:
\begin{align}
(1-x-2y)&J_n^{(a,b,c)}(x,y)
    =(1-x-2y)J_{n-k}^{(a,b+c+2k+1)}(x)\, (1-x)^k \,J_k^{(b,c)}\left( \frac{y}{1-x}\right)\nonumber\\
& = J_{n-k}^{(a,b+c+2k+1}(x) \,(1-x)^{k+1} \,\left( 1- \frac{2y}{1-x}\right)\, J_k^{(b,c)}\left(\frac{y}{1-x}\right)\nonumber\\
&=
A_{k}^{(1)} \,J_{n-k}^{(a,b+c+2k+1)}(x) \,(1-x)^{k+1}\,
 J_{k+1}^{(b,c)}\left( \frac{y}{1-x}\right)\nonumber\\
&\quad+A_{k}^{(2)}\,(1-x) \, J_{n-k}^{(a,b+c+2k+1)}(x) \,(1-x)^{k}\,
J_{k}^{(b,c)}\left( \frac{y}{1-x}\right)\nonumber\\
&\quad +A_{k}^{(3)}\,(1-x)^2 \,J_{n-k}^{(a,b+c+2k+1)}(x) \,(1-x)^{k-1}\,
J_{k-1}^{(b,c)}\left(\frac{y}{1-x}\right),
\label{RR2proof:01}
\end{align}
where we use the recurrence relation (\ref{RR:univariate}) for the univariate Jacobi polynomials.
First, the term in the fourth line of (\ref{RR2proof:01}) can be expressed using $J_{n,k}^{(a,b,c)}$, and by applying the recurrence relation (\ref{RR1}) for the bivariate Jacobi polynomials, it can be rewritten as a linear combination of the bivariate polynomials $J_{n+1,k}^{(a,b,c)}$, $J_{n,k}^{(a,b,c)}$, and $J_{n-1,k}^{(a,b,c)}$.
Next, the $J_{n-k}^{(a,b+c+2k+1)}$ appearing in the third line can be expanded using  (\ref{uniJacobi:GTb}) in terms of  $J_{n-k}^{(a, b+c+2k+3)}$, $J_{n-k-1}^{(a, b+c+2k+3)}$, and $J_{n-k-2}^{(a, b+c+2k+3)}$. As a result, the third line can be expressed as a sum of the bivariate polynomials $J_{n+1,k+1}^{(a,b,c)}$, $J_{n,k+1}^{(a,b,c)}$, and $J_{n-1,k+1}^{(a,b,c)}$.
Similarly, $(1-x)^2 \,J_{n-k}^{(a,b+c+2k+1)}$ in the last line of (\ref{RR2proof:01}) can be expanded using (\ref{uniJacobi:CTb}) in terms of the bivariate polynomials $J_{n-k+2}^{(a, b+c+2k-1)}$, $J_{n-k+1}^{(a, b+c+2k-1)}$, and $J_{n-k}^{(a, b+c+2k-1)}$. Then the last line can be expressed as a sum of $J_{n+1,k-1}^{(a,b,c)}$, $J_{n,k-1}^{(a,b,c)}$, and $J_{n-1,k-1}^{(a,b,c)}$. 
Finally, by reorganizing the coefficients of $J_{n+i,\,k+j}^{(a,b,c)}(x,y)$ for $-1 \le i, j \le 1$ as obtained above, we arrive at the recurrence relation $(1-x-2y)J_n^{(a,b,c)}(x,y) = (2 \Hat{X}_3 + \hat{X}_1 -I)J_n^{(a,b,c)}(x,y)$, which leads to (\ref{RR2}).\end{proof}

\section{The rank two Jacobi algebra from the variable realization}
This central section aims to offer a presentation of the rank two Jacobi algebra. It is hence devoted to the determination of generators and defining relations that will be abstracted from the model provided by the two variable Jacobi polynomials. Beginning with the commutators of the bispectral operators, this will be achieved by pursuing the cascade of commutation relations between the resulting operators until closure in terms of quadratic expressions in a set of generators is attained. The number of these relations is somewhat large and will be reported in a number of subsections corresponding to the step-wise process that leads to them. 

\subsection{Commutators of the bispectral operators: the operators $M_1$, $M_3$, $N_1$ and $N_3$}

The first step is to compute the commutators of the bispectral operators in the variable representation which are:
\begin{itemize}
    \item the two differential operators:
    \begin{equation}
        L \qquad \text{and} \qquad L_1;\label{dif}
    \end{equation}
    \item the operators associated to the two recurrence relations that we shall denote by:
    \begin{equation}
        X_1=x \qquad \text{and} \qquad X_3=1-x-y.\label{pos}
    \end{equation}
\end{itemize}

\noindent Since $L$ and $L_1$ are simultaneously diagonal on the polynomials $J_{n,k}^{(a,b,c)}(x,y)$, they commute and it is manifest that the same is true of $X_1$ and $X_3$, that is:
\begin{equation}
    [L, L_1]=0 \qquad \text{and} \qquad [X_1,X_3]=0.
\end{equation}
The commutators of the bispectral operators \eqref{dif} with the \eqref{pos} ones yield the new generators $N_1, N_3, M_1, M_3$ that are given by:
\begin{align}
    [L,X_1]&=N_1= 2x(1-x)\partial_x - 2xy \partial_y +a+1 -(a+b+c+3)x, \label{Nx}\\
    [L,X_3]&=N_3 
     =2 (x+y-1)(x\partial_x + y \partial_y) + c+1 + (a+b+c+3)(x+y-1),\label{Ny}\\
    [L_1,X_1]&=M_1=0,\\
    [L_1,X_3]&=M_3=2y(x+y-1)\partial_y + (b+1)(x+y-1) + (c+1)y.\label{My}
\end{align}
\subsection{Supplementary relations}
This then calls for the evaluation of the commutators of these additional generators with the bispectral operators. The results are found below. 

\subsubsection{The commutators of $N_1$ with the bispectral operators}
\begin{align}
    [N_1,X_1]&=-2X_1^2+2 X_1,\label{XXL}\\
    [N_1,X_3]&=-2X_1 X_3,\\
    [\,N_1,L\;]&=2\{X_1,L\}-2L+2L_1-(a+b+c+1)((a+b+c+3)X_1 -(a+1)I),\label{LLX}\\
    [N_1,L_1]&=0. 
\end{align}
Note: it is observed by comparing \eqref{XXL} and \eqref{LLX} with \eqref{rank1} that $L$ and $X_1$ generate a rank one Jacobi subalgebra where $L_1$ is central since $[L,L_1]=0$ and $[X_1,L_1]=0$. Specifically, write this central element as
\begin{equation}
    L_1=\ell (b+c-\ell +1).
\end{equation}
Upon taking
\begin{equation}
    K_1=L-\ell(a+b+c-\ell+2) \qquad \text{and } \qquad K_2=X_1,
\end{equation}
with
\begin{equation}
    \alpha=a \qquad \text{and} \qquad \beta=b+c-2\ell +1,
\end{equation}
it is readily seen that the relations \eqref{rank1} translate into \eqref{XXL} and \eqref{LLX}.

\subsubsection{The commutators of $N_3$ with the bispectral operators}
\begin{align}
    [N_3,X_1]&=-2X_1 X_3,\\
    [N_3,X_3]&=-2X_3^2+2X_3, \label{NyY}\\
    [\,N_3,L\;]&=2\{ X_3,L\}-2L+2 L_3
    -(a+b+c+1)((a+b+c+3)X_3 - (c+1)I), \label{NyL}\\
    [N_3,L_1]&=\{X_1-I, L- L_3\}+\{X_3,L+L_1\}
\nonumber\\& \qquad -(a+b+c+1)\left((c+1)(X_1+X_3-I)+(b+1)X_3\right)\label{NyLy}.
\end{align}
Note: from this series of commutators we find that $L$ and $X_3$ generate a rank one Jacobi subalgebra with $L_3$ central ($[L,L_3]=0, [X_3,L_3]=0$). To confirm this, write the central element in the form:
\begin{equation}
    L_3=\ell (a+b-\ell+1).
\end{equation}
Set then,
\begin{equation}
    K_1=L-\ell (a+b+c-\ell +2) \qquad \text{and} \qquad K_2=X_3,
\end{equation}
with
\begin{equation}
    \alpha =c \qquad \text{and} \qquad \beta=a+b-2\ell+1.
\end{equation}
Injecting these identifications in the relations \eqref{rank1} that define the rank one Jacobi algebra gives the  commutation relations \eqref{NyY} and \eqref{NyL}.

\subsubsection{The commutators of $M_3$ with the bispectral operators}
\begin{align}
    [M_3,X_1]&=0,\\
    [M_3,X_3]&=-2X_3^2+2 (I-X_1)X_3,\label{MyY}\\ 
    [\,M_3,L\;]&= \{X_1-I,L-L_3\}+\{X_3, L+L_1\}\nonumber\\& \qquad
    -(a+b+c+1)\left((c+1)(X_1+X_3-I)+(b+1)X_3\right)=[N_3,L_1], 
\label{MyL}
    \\
    [M_3,L_1]&=2\{X_3 ,L_1\}+\{X_1,L_1\}-2L_1 -(b+c)\left((b+c+2)X_3-(c+1)(I-X_1) \right).
    \label{MyLy}
\end{align}
\noindent
Note: here we see that $L_1$ and $X_3$ also generate a rank one Jacobi subalgebra with $X_1$ central since $[L_1,X_1]=0$ and $[X_3,X_1]=0$. 

For $X_1 \neq 1$, this is checked as follows. Setting
\begin{equation}
    K_1=L_1 \qquad \text{and} \qquad K_2 = \frac{X_3}{X_1 -1}\;+1 \qquad \text{with} \qquad \alpha = b, \qquad \beta =c,
\end{equation}
it is immediate to see that the relations \eqref{rank1} are transformed into \eqref{MyY} and \eqref{MyLy}.

If $X_1=1$, the subalgebra that $L_1$ and $X_3$ lead to is a degeneration of the rank one Jacobi algebra which is obtained from \eqref{rank1} through a contraction achieved from scaling $K_2 \rightarrow \tau K_2$ and taking the limit $\tau \rightarrow \infty$. This will yield the commutation relations that are obtained from \eqref{MyY} and \eqref{MyLy} when $X_1=1$ upon keeping the identification $\alpha=b,\; \beta=c$ and that define actually a Lie algebra as shown in the Appendix 1 of \cite{granovskii1992mutual}.

\subsubsection{Upshot}
A look at all the relations gathered in this subsection shows that the triple commutators among 
the operators $L, L_1, X_1, X_3$
close under symmetric quadratic expressions in the elements of this set
except for the presence of $L_3$ in the right hand sides of equations \eqref{NyL}, \eqref{NyLy} and \eqref{MyL} giving the expressions for $[N_3,L]$, $[N_3,L_1]$ and $[M_3,L]$ respectively. 

\subsection{Towards closure: adding $L_3$ as generator}
This last observation forces the inclusion of $L_3$ in the set of generators that will now be $\{L, L_1, L_3, X_1, X_3\}$.
Computing the commutators of this new generator with the former ones leads to the following additional operators:
\begin{align}
    [L_3, X_1]=&J_1=2xy(\partial_x - \partial_y) +((a+1)y - (b+1)x)I   =-M_3+N_1+N_3, \label{Kx}\\
    [L_3,X_3]=&J_3= 0, \label{Ky}\\
    [L_3,L_1]=&-G_{13}=\{M_3,L\}-\frac12 \{N_3,L+L_1\}-\frac12 \{N_1,L-L_3\} \nonumber\\
    &-\frac{1}{2}(a+b+c+1)\left((a+b+c+3)M_3 -(b+c+2)N_3-(c+1)N_1\right) . 
    \label{LzLy}
\end{align}
We will remember that $[L_3,L]=0$
and since $X_3 = 1 - x - y$ commutes with $L_3$, it follows that $J_3 = 0$.

It is then seen that the additional non-trivial triple commutators involving at least one $L_3$ and the other fundamental generators are symmetric quadratic combinations of the elements of the extended set of generators. These triple commutators can be put in two categories those that involve $L_3$ once and those where $L_3$ appear twice. In the first category we have
\begin{itemize}
   \item The commutators of $L_3$ with one of $\{N_1, N_3, M_3\}$:
\begin{align}
    [L_3,N_1]&=[L_3,[L,X_1]]    \nonumber\\
     & =- \{X_1,L+L_3\}-\{X_3-I,L-L_1\}\nonumber\\
     & \quad +(a+b+c+1)\left((a+1)(X_1+X_3-I)+(b+1)X_1\right), \\
    [L_3,N_3]&=[L_3,[L,X_3]] =0, \\
    [L_3,M_3]&=[L_3,[L_1,X_3]]
    \nonumber \\
     &= -\{X_1,L-L_3\} - \{X_3-I,L-L_1-L_3\} \nonumber \\
     & \quad +(c+1) \left((a+1)(X_1+X_3-I) + (b+1)X_1 \right). 
\end{align}
   \item The commutators of one of the bispectral operators $L, L_1, X_1, X_3$ with one of the set $\{J_1, G_{13}\}$ 
\end{itemize}
\begin{align}
    [L,J_1]=&[L,[L_3,X_1]]=-\{X_1,L+L_3\} - \{X_3-I,L-L_1\}
\nonumber\\
     & + (a+b+c+1) \left((a+1)(X_1+X_3-I) + (b+1)X_1\right)
    =[L_3,N_1],\\
    [L_1,J_1]=&[L_1,[L_3,X_1]]=
 - \{X_1-I,L-L_1-L_3\} - \{X_3,L-L_1\} 
\nonumber\\
    & \quad + (a+1)\left((b+1) X_3 + (c+1)(X_1+X_3-I)\right),\\
    [X_1,J_1]=&[X_1,[L_3,X_1]]=2X_1(X_1+X_3-I),\label{eq:X1J1}\\
    [X_3,J_1]=&[X_3,[L_3,X_1]]= 0  =[X_1,J_3],\\
    [X_1,G_{13}]=&[X_1,[L_1,L_3]]=
    \{X_1-I,L-L_1-L_3\}+\{X_3,L-L_1\}
\nonumber\\
    &-(a+1)\left((b+1)X_3+(c+1)(X_1+X_3-I)\right)=-[L_1,J_1],\\
    [X_3,G_{13}]=&[X_3,[L_1,L_3]]=
-\{X_1,L-L_3\} - \{X_3-I,L-L_1-L_3\}
\nonumber \\
    & 
+(c+1)\left((a+1)(X_1+X_3-I)+(b+1)X_1\right)
=[L_3,M_3].
\end{align}
The commutator $[L_1,G_{13}]=[L_1,[L_1,L_3]]$ that belongs to this list will be treated separately below.

In the second category of triple commutators i.e. of those that involve two $L_3$ we have:
\begin{itemize}
    \item The commutators of $L_3$ with $J_1$:
  \begin{align}
[L_3,J_1]&= [L_3,[L_3,X_1]]=-\{2X_1+X_3-I,L_3\} \nonumber\\ 
 &+(a+b)\left((a+1)(X_1+X_3-I)+(b+1)X_1\right).\label{eq:L3J1}
  \end{align}  
  \item The commutator $[L_3, G_{13}]=[L_3,[L_1,L_3]]$ that will also be treated below.
\end{itemize}

Note: At this point, we can identify another rank one Jacobi subalgebra generated in this case by $L_3$ and $X_1$ with $X_3$ central ($[L_3, X_3]=0, [X_1,X_3]=0)$. 

If $X_3\neq 1$, one takes
\begin{equation}
    K_1=L_3 \qquad K_2=\frac{X_1}{X_3-1}\; +1 \qquad \text{with} \qquad \alpha=b, \; \beta=a, \label{JacsubL3X1}
\end{equation}
to find that the relations \eqref{rank1} transform into the equations \eqref{eq:X1J1} and \eqref{eq:L3J1}.

If $X_3=1$, $L_3$ and $X_1$ realize the same contraction of the Jacobi algebra as $L_1$ and $X_3$ except that here the identification of the parameters is the one given in \eqref{JacsubL3X1}.

\subsection{Remarks on relations between the triple commutators}
The ensemble of triple commutators was formed so as to be exhaustive through a systematic enumeration (barring manifestly trivial cases such as $M_1=0$ and $J_3=0$). There are however relations between some of them. First, the commutators involving $J_1$ can be obtained in terms of those featuring $N_1$, $N_3$ and $M_3$ in view of the linear expressions \eqref{Kx} giving $J_1$ in terms of the latter set of operators. Second, the Jacobi identity implies relations between some of the triple commutators listed. They are the following.
 Applying $[X,[Y,Z]]+[Y,[Z,X]]+[Z,[X,Y]]=0$, one finds for instance \
\begin{equation}
    [L_1, J_1]=-[X_1,G_{13}] \quad \text{since} \quad [L_1,X_1]=0,
\end{equation}
or,
\begin{equation}
    [L_3, N_1]=[L,J_1] \quad \text{using} \quad [L_3,L]=0.
\end{equation}
The other cases are
\begin{align}
&[N_1,X_3]=[N_3,X_1],\\
&[M_3,L]=[N_3,L_1,]\\
    &[L_3,M_3]=[X_3,G_{13}],\\
    &[X_3,J_1]=0,\\ 
    &[L_3,N_3]=0. 
\end{align}
 In order to allow the tracking of independent relations, these last relations stemming from the Jacobi identity have been recorded in the preceding list whenever a triple commutator can be expressed in terms of commutators that have appeared before in the sequence.

If we set $X_2 = y$, call upon $L_2$, and further define $\alpha_1 = a$, $\alpha_2 = b$, and $\alpha_3 = c$, we can verify that the relations given above exhibit a symmetry: they still hold, for example, under the interchange of the indices 2 and 3. That is, the relations appear to hold for any choice of index pairs among (1,2), (1,3), and (2,3).

\subsection{The rank one Racah subalgebra}
The determination of the triple commutators 
$[L_1,G_{13}]=[L_1,[L_1,L_3]]$ and $[G_{13},L_3]=[L_3,[L_3,L_1]]$  
was left aside. These must of course be included in the defining relations of the rank two Jacobi algebra; focusing on them will make us realize that in addition to the rank one Jacobi subalgebras that we have identified so far, the rank two Jacobi algebra further admits as subalgebra, a rank one Racah algebra generated by $L_1$ and $L_3$. 

Recall that the generic presentation of the rank one Racah algebra \cite{genest2014racah} involves the following relations between the two generators $\Hat{K}_1$ and $\Hat{K}_2$:
\begin{align} \label{Racahgen}
    [\Hat{K}_1,[\Hat{K}_1,\Hat{K}_2]]=&-a_1\Hat{K}_1^2-a_2\{\Hat{K}_1,\Hat{K}_2\}-c_2\Hat{K}_2-d\Hat{K}_1-e_2, \\
       [\Hat{K}_2,[\Hat{K}_2,\Hat{K}_1]]=&-a_2\Hat{K}_2^2-a_1\{\Hat{K}_1,\Hat{K}_2\}-c_1\Hat{K}_1-d\Hat{K}_2-e_1,
\end{align}
where the parameters $a_1, a_2, c_1, c_2, d, e_1, e_2$ are real. It can be checked that the rank one Jacobi algebra presented in \eqref{rank1} is indeed a special case of \eqref{Racahgen} with certain assignments of the parameters and in particular $a_1=0$. Now, if this algebra genuinely describes the family of Racah polynomials and not one of their descendants neither $a_1$ or $a_2$ can be zero, i.e. $a_1a_2\neq0$. When this is so, the following scalings: $\Hat{K}_1=\frac{1}{2}a_2K_1$ and $\Hat{K}_2=\frac{1}{2}a_1K_2$ cast the defining relations of the rank one Racah algebra in the following form:
\begin{align}
    [K_1,[K_1,K_2]]=&-2\{K_1,K_2\}-2K_1^2+\alpha K_1+\beta K_2+\gamma, \label{Rac1}\\
    [K_2,[K_2,K_1]]=&-2\{K_1,K_2\}-2K_2^2+\alpha K_2+\delta K_1+\epsilon.\label{Rac2}
\end{align}
Note that the shifts $K_1 \rightarrow K_1+\frac{1}{4}\beta$ and $K_2 \rightarrow K_2+\frac{1}{4} \delta$ can eliminate, if desired, the redundant parameters $\beta$ and $\delta$ in the above relations while preserving the equality of the coefficient of $K_1$ on the rhs of $  [K_1,[K_1,K_2]]$ and of the one of $K_2$ on the rhs of $[K_2,[K_2,K_1]]$; this yields the usual standard presentation that can be found in \cite{genest2014racah} for example. We may finally remark that the Jacobi algebra \eqref{rank1} is attained from the presentation \eqref{Rac1} and \eqref{Rac2} by a contraction process where the generator $K_2$ and some parameters are rescaled according to $K_2 \rightarrow \tau K_2, \;
\alpha \rightarrow \tau \alpha, \; \gamma \to \tau \gamma$ and where one subsequently lets $\tau \rightarrow \infty$ while taking $\alpha =2$.

With the help of \eqref{LzLy}, the following triple commutators can be computed to find:
    \begin{align}
    [L_1,[L_1,L_3]]&=-2\{L_1,L_3\}-2L_1^2+2L_1L \nonumber\\&
    -(b+c)(b+1)(L-L_1-L_3)
    + (b+c)(c+1)L_3+(b-c)(a+1)L_1, \label{LyLz1}\\
    [L_3,[L_3,L_1]]&=-2\{L_1,L_3\}-2L_3^2+2L_3L  \nonumber\\&
    -(a+b)(b+1)(L-L_1-L_3)
    + (a+b)(a+1)L_1+(b-a)(c+1)L_3,\label{LyLz2}
\end{align}
where $[L,L_1]=[L,L_3]=0$.

Comparing the equations \eqref{LyLz1} and \eqref{LyLz2} with the relations \eqref{Rac1} and \eqref{Rac2}, we observe under the identification $K_1 \rightarrow L_1$ and $K_2 \rightarrow L_3$ that the operators $L_1$ and $L_3$ realize together a central extension of the rank one Racah algebra where the parameter $\alpha$ is in particular identified to:
\begin{equation}
    \alpha=(b+c)(b+1)+(b-c)(a+1)+2L=(a+b)(b+1)+(b-a)(c+1)+2L.
\end{equation}

Since $L_1$ and $L_3$ commute with $L$, it follows that the Racah algebra of rank one is the symmetry algebra of $L$ and that the two-variable Jacobi polynomials $J_{n,k}^{(a,b,c)}(x,y)$ with $n$ fixed transform among themselves under the action of its generators (see next section).

\subsection{Wrap up}
Contained in section 3 are all the elements that define the rank two Jacobi algebra in terms of generators and relations. To sum-up and guide the reader through the myriad of formulas, we conclude this section by recapping the salient features of the presentation.
\subsubsection{Generators} The set of generators consists of the following elements:
\begin{equation}
\{L, L_1 , L_3 , X_1, X_3 \}.     
\end{equation}
\subsubsection{The primary commutators} These are the commutators of all the generators among themselves:
\begin{equation}
    [L, L_1 ] = [L, L_3 ] = [X_1, X_3 ] = 0, \quad[L, X_1] = N_1, \quad[L, X_3 ] = N_3 ; 
    \end{equation}
    \begin{equation}
[L_1 , L_3 ] = G_{13}, \quad[L_1 , X_1] = M_1 = 0, \quad[L_1 , X_3 ] = M_3 , \quad[L_3 , X_1] = J_1,\quad [L_3 , X_3 ] = J_3 =0 . 
\end{equation}
This forms a list of 5 elements:
\begin{equation}\label{pri}
    \{N_1, N_3 , M_3 , G_{13} , J_1\}.
\end{equation}
\subsubsection{The triple commutators} The defining relations consists in the expression of all the triple
commutators of the generators which amount to the commutators of the fundamental generators
with the 5 primary commutators listed in \eqref{pri}. These are expressed in terms of symmetric
quadratic expressions in the generators. They will be found in the subsections (3.2), (3.3) and (3.4). Modulo the equalities and the relations between these triple commutators and in particular those implied by the Jacobi identity, they form the defining relations of the rank two Jacobi identities.

\subsubsection{Subalgebras} In the course of determining the defining relations, significant subalgebras were identified. Note that the centralizers of each element of the generating set of the rank two Jacobi algebra is two-dimensional. For each element, these are:
\begin{align}
    L_1 \qquad :& \qquad \{L, X_1\} \\ \label{sa1}
    X_1 \qquad :& \qquad \{L_1, X_3\} \\
  X_3 \qquad :& \qquad \{X_1, L_3\} \\
   L_3 \qquad :& \qquad \{X_3,L\}\\
    L \qquad :& \qquad \{L_3,L_1\}.\label{sa5}
\end{align}
The first four centralizers form rank one Jacobi algebras while the last one realizes the rank one Racah algebra. A feature of this ordered list is that the second generator of a chosen subalgebra is central in the following one and the property is cyclic.

Therefore, these subalgebras can be neatly displayed in a pentagon diagram as in Figure 1 below. This is achieved by associating cyclically in an ordered fashion to a vertex the second generator of a certain centralizing set and the first generator of the following pair. These two generators commute by design. The generators of the subalgebras  \eqref{sa1} - \eqref{sa5} are hence attached to each edge of the pentagon. This offers a meaningful way of summarizing the subalgebra structure of the rank 2 Jacobi algebra that was uncovered.

\begin{figure}[htbp]
\begin{center}
\begin{minipage}[t]{.9\linewidth}
\begin{center}
\begin{tikzpicture}[scale=1.3]
   
    \def\R{2cm}
    \def\Rt{2.1cm}
    \def\Rtt{1.85cm}

    \foreach \i in {1,...,5} {
        \coordinate (P\i) at ({90 + (\i-1)*360/5}:\R);
    }
    \draw (P2) -- (P1) -- (P5) -- (P4) -- (P3);
    \draw[dashed] (P2) -- (P3);

\node at ({36 +90+ (0)*360/5}:\Rtt) {$N_1$};
\node at ({36 +90+ (1)*360/5}:\Rtt) {$G_{13}$};
\node at ({36 +90+ (2)*360/5}:\Rtt) {$N_3$};
\node at ({36 +90+ (3)*360/5}:\Rtt) {$J_1$};
\node at ({36 +90+ (4)*360/5}:\Rtt) {$M_3$};

\node at ({90 + (+0.12)*360/5}:\Rt) {$X_1$};
\node at ({90 + (-0.12)*360/5}:\Rt) {$L_1$};
\node at ({90 + (-0.12+1)*360/5}:\Rt) {$L$};
\node at ({90 + (+0.12+1)*360/5}:\Rt) {$L_1$};
\node at ({90 + (-0.12-1)*360/5}:\Rt) {$X_1$};
\node at ({90 + (+0.12-1)*360/5}:\Rt) {$X_3$};
\node at ({90 + (-0.12-2)*360/5}:\Rt) {$X_3$};
\node at ({90 + (+0.12-2)*360/5}:\Rt) {$L_3$};
\node at ({90 + (-0.12+2)*360/5}:\Rt) {$L_3$};
\node at ({90 + (+0.12+2)*360/5}:\Rt) {$L$};

     \foreach \i in {1,...,5} {
        \draw[fill]  (P\i) circle (0.08); ;
     }

\end{tikzpicture}
\caption{Rules: Two generators around the same vertex commute; two generators at the boundaries of the same solid line generate a Jacobi algebra; two generators at the boundaries of the same dashed edge generate a Racah algebra; the generator at the middle of an edge corresponds to the commutator of the generators at the boundaries of this edge.   \label{fig:tr}}
\end{center}
\end{minipage}
\end{center}
\end{figure}
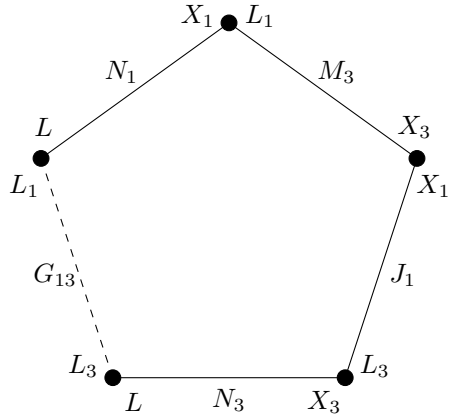

%%%%%%% 

\subsection{Connection to the rank two Racah algebra} How the rank two Jacobi algebra that has been defined relates to a contraction of the rank two Racah algebras $R(4)$ will now be described briefly.

The algebra $R(4)$  has been introduced and studied in different contexts \cite{GI2010,post2015,de2017higher,Gaboriaud2019,Debie2020,LMZ2020,crampe2021racah,crampe2023representations, post2024racah,Crampe2024} and there exist numerous presentations of this algebra. The remarks that follow refer to the one given in  \cite{crampe2023representations} in a standardized form.
 $R(4)$ has five generators denoted by $C_{12}, C_{23}, C_{34}, C_{123}, C_{234}$ and five central elements $C_1,C_2,C_3,C_4,C_{1234}$. Its defining relations are collected in the Appendix A of \cite{crampe2023representations}.

Consider the contraction $\cJ(4)$ of $R(4)$ obtained by performing the following rescalings
\begin{equation}
    C_{12} \rightarrow \tau C_{12}, \quad  C_{123} \rightarrow \tau C_{123}, \quad  C_{1234} \rightarrow \tau C_{1234},
\end{equation}
while keeping the other generators unchanged and retaining in the relations only the leading terms in $\tau$ as $\tau \rightarrow \infty$.

A straightforward computations shows that the following assignments (up to constant shifts)
\begin{align}
  &  C_{12} \rightarrow X_3,\quad C_{23} \rightarrow -L_1+\frac14(b+c)(b+c+2),\quad C_{34} \rightarrow -L_3+\frac14(a+b)(a+b+2),\nonumber\\
   &C_{123} \rightarrow 1-X_1,\quad C_{234} \rightarrow -L+\frac14(a+b+c+3)(a+b+c+1),\label{eq:RelJ1}
\end{align}
to the operators that lead to the rank two Jacobi algebra
with
\begin{align}
  C_1=0 ,\quad C_2=\frac14(c^2-1) ,\quad C_3=\frac14(b^2-1) ,\quad C_4=\frac14(a^2-1) ,\quad   C_{1234}=1,\label{eq:RelJ2}
\end{align}
provide a realization of $\cJ(4)$.

\section{Representation on the bivariate Jacobi polynomial basis or the degree realization}
We consider in this section the dual realization of the rank two Jacobi algebra in terms of operators involving the degrees $n$ and $k$ of the two variable Jacobi polynomials $J_{n,k}^{(a,b,c)}$. This amounts to providing the representation of the rank two Jacobi algebra on the space with basis vectors realized by the two variable Jacobi polynomials. When taken in conjunction with the variable realization of section 3, the determination of the primary commutators of the generators in this degree model will be seen to lead to structure relations for the two variable polynomials that will be presented.

\subsection{The generators in the degree representation} 
The two variable Jacobi algebra has five generators that were given in the variable representation by the operators $L, L_1, L_3, X_1, X_3$. Their representations in the degree model shall be denoted by $\Hat{L}, \Hat{L}_1, \Hat{L}_3, \Hat{X}_1, \Hat{X}_3 $. The bispectral operators are readily provided by the properties of the two variable Jacobi polynomials. The difference analog $\Hat{L}_3$ of $L_3$ is the one that needs further attention. Here they are:

\subsubsection{The bispectral operators (in the degree model)}
\begin{itemize}
    \item The difference operators:
    \begin{equation}
        \Hat{X}_1, \qquad \Hat{X}_3
    \end{equation}
   which are respectively given by \eqref{M2} and \eqref{M}, and define the recurrence relations \eqref{RR1} and \eqref{RR2}.
   \item The multiplication operators by the eigenvalues of the differential operators $L$ and $L_1$:
   \begin{equation}
       \Hat{L}=-n\,(n+a+b+c+2), \qquad\Hat{L}_1=-k\,(k+b+c+1).
   \end{equation}
\end{itemize}

\subsubsection{$\Hat{L}_3$}

\begin{align}
    \Hat{L}_3&=\frac{ \left(k+b  \right) \left(k+c  \right)\left(n-k+1\right) \left(n +k +b +c +1\right)}{\left(2k+b +c  \right) \left(2 k +b +c +1\right)}  
    T_{-}
    \nonumber \\&
    +\frac{\left(k +1\right) \left(k +b +c+1\right) \left(n -k+a \right) \left(n +k +a +b +c +2\right) }{\left(2 k +b +c +1\right) \left(2 k +b +c +2\right)}
    T_{+}
    \nonumber \\&
    +\biggl((k-n)(n-k+a+b+1) -\dfrac{k(k+c)(n-k+1) (n-k+a+1)}{2k+b+c} 
    \nonumber \\&\qquad
    + \dfrac{(k+1)(k+c+1)(n-k)(n-k+a)}{2k+b+c+2} \biggr) I .
\label{HLz}
\end{align}
\begin{proof}
The determination of this difference operator was obtained in \cite{iliev2018symmetry} in different notations, as part of the construction of the representation of the symmetry algebra of $L$ in the two variable Jacobi polynomial basis. This expression for $\Hat{L}_3$ can be straightforwardly obtained using the differential properties collected in Appendix B. Consider the operators $s_3$ and $s_3^*$ given in \eqref{diffop}. The result follows from the observation that $L_3=(s_3^* -1)s_3+b(a+1)$, see \eqref{Lzs}, and from the actions of $s_3$ and $s_3^*$ respectively given by \eqref{acts3} and \eqref{acts3*}. Alternatively, \eqref{HLz} can be obtained from the ``hatted'' version of equation \eqref{NyL} that allows to obtain $\Hat{L}_3$ from the knowledge of $\Hat{L}$, $\Hat{L}_1$, $\Hat{X}_3$, $[[\Hat{L}, \Hat{X}_3], \Hat{L}]$ and $\{\Hat{L},\Hat{X}_3\}$ all following from the formulas for the bispectral operators in the degree representation.
\end{proof}
\subsection{Observations}
Let us denote by $O$ an operator from the set $\{L, L_1, L_3, X_1, X_3\}$ of the variable representation and by $\Hat{O}$ the corresponding operator in the degree representation. Given that the difference operators provide the actions of the differential ones on the polynomial basis:
\begin{equation}
    OJ_{n,k}^{(a,b,c)}(x,y)=\Hat{O}J_{n,k}^{(a,b,c)}(x,y), \label{repres}
\end{equation}
it follows that the operators $\Hat{O}$ and $\Hat{O}'$ will satisfy the same commutation relations as $O$ and $O'$. For verification purposes, we have checked that fact to ensure the correctness of all the formulas.

These operators $\hat{O}$ hence provide a representation on a basis that can be denoted by $\{|n,k; a, b,c\rangle\}$ which is realized by the two-variable Jacobi polynomials: $\langle x,y|n,k;a,b,c\rangle=J_{n,k}^{(a,b,c)}(x,y)$. With $\mathcal{O}$ denoting the representation in this basis of the corresponding abstract generator, one has then $\langle x,y|\mathcal{O}|n,k;a,b,c\rangle=OJ_{n,k}^{(a,b,c)}(x,y)$ and keeping with the calligraphic notation, we can write down the following representation of the rank two Jacobi algebra:
\begin{align}
    \mathcal{L}&|n,k;a,b,c\rangle=-n(n+a+b+c+2)|n,k;a,b,c\rangle,\\
    \mathcal{L}_1&|n,k;a,b,c\rangle=-k(k+b+c+1)|n,k;a,b,c\rangle,\\
    \mathcal{X}_1&|n,k;a,b,c\rangle=\Hat{X}_1|n,k;a,b,c\rangle,\\
    \mathcal{X}_3 &|n,k;a,b,c\rangle=\Hat{X}_3|n,k;a,b,c\rangle,\\
  \mathcal{L}_3& |n,k;a,b,c\rangle=\Hat{L}_3|n,k;a,b,c\rangle,
\end{align}
with 
\begin{equation}
    T_{\pm}|n,k;a,b,c\rangle=|n\pm 1,k;a,b,c\rangle, \qquad S_{\pm}|n,k;a,b,c\rangle=|n,k\pm 1;a,b,c\rangle.
\end{equation}
\subsection{The primary commutators in the degree representations: the operators $\Hat{N}_1,\Hat{N}_3, \Hat{M}_3,\\
\Hat{K}_1$}
With the expression of all the generators in the degree representation in hand, it is simply a matter of computing the commutators of these difference operators to obtain the analogs of $N_1, N_3, M_3, K_1$ and here they are:
\begin{align}
    [\Hat{L},\Hat{X}_1 ]=&\Hat{N}_1=
\frac{\left(n-k+1\right) \left(n +k +a +b +c +2\right)}{2 n +a +b +c +2} S_{+}-
\frac{\left(n -k + a \right) \left(n +k +b +c +1\right)}{2 n +a +b +c +2} S_{-} 
,\label{HNx}\\
%%%
    [\Hat{L},\Hat{X}_3 ]=&\Hat{N}_3  \nonumber\\
  =& -\frac12 \Hat{N}_1 
  -\frac{\left(k +1\right) \left(k +b +c +1\right) \left(n +k +a +b +c +2\right) \left(n +k +a +b +c +3\right) }{ \left(2 k +b +c +1\right) \left(2 k +b +c +2\right)\left(2 n +a +b +c +2\right)}
    S_{+}T_{+} \nonumber\\
   & +\frac{\left(b -c \right) \left(b +c \right)\left(n-k+1\right) \left(n +k +a +b +c +2\right)  }{ 2\left(2 k +b +c \right) \left(2 k +b +c +2\right) \left(2 n +a +b +c +2\right) }
     S_{+} \nonumber\\
   &-\frac{\left(k +b \right) \left(k +c\right) \left(n-k+1\right) \left(n-k+2\right) }{\left(2k+b +c\right) \left(2k+b +c+1 \right) \left(2 n +a +b +c +2\right)}
    S_{+}T_{-}  \nonumber\\
   & +\frac{\left(k +1\right)\left(k +b +c +1\right) \left(n -k+a -1\right)\left(n -k+a \right)   }{ \left(2k+b +c +1\right) \left(2k+b +c +2\right)\left(2n+a +b +c +2\right)}
    S_{-}T_{+} \nonumber\\
    &-\frac{\left(b -c \right) \left(b +c \right) \left(n -k+a  \right) \left(n +k +b +c +1\right) }{2 \left(2 k+b +c \right) \left(2k+b +c +2\right)\left(2 n +a +b +c +2\right) }
    S_{-} \nonumber\\ 
    &+\frac{\left(k+b \right) \left(k+c \right) \left(n +k +b +c \right) \left(n +k+b +c  +1\right) }{\left(2 k +b +c \right) \left(2 k+b +c+1  \right)\left(2 n +a +b +c +2\right) } 
    S_{-}T_{-}    
    ,\label{HNy}\\
    [\Hat{L}_1,\Hat{X}_3 ]=&\Hat{M}_1=0,\\
    [\Hat{L}_1,\Hat{X}_3 ]=&\Hat{M}_3 
    \nonumber \\=&
    -\frac{\left(k +1\right) \left(k +b +c +1\right) \left(n+k+a+b+c+2\right) \left(n+k+a+b+c+3\right) }{\left(2k+b+c+1\right) \left(2n+a+b+c+2\right) \left(2n+a+b+c+3\right) }
    S_{+}T_{+} 
    \nonumber \\&
    +\frac{\left(k+b \right) \left(k+c \right) \left(n-k+1\right) \left(n-k+2\right) }{\left(2k+b +c+ 1\right) \left(2n+a +b +c +2\right) \left(2n+a+b +c +3\right)}
    S_{+}T_{-} 
    \nonumber \\&
    -\frac{2 (k+1)(k+b+c+1)(n-k+a)(n+k+a+b+c+2)}{\left(2k+b +c+ 1\right) \left(2n+a +b +c +1\right) \left(2n+a+b +c +3\right)}
    T_{+} 
    \nonumber\\&
    +\frac{2(k+b)(k+c)(n-k+1)(n+k+b+c+1)}{\left(2k+b +c+ 1\right) \left(2n+a +b +c +1\right) \left(2n+a+b +c +3\right)}
    T_{-} 
    \nonumber \\&
    -\frac{(k+1)(k+b+c+1)(n-k+a-1)(n-k+a)}{\left(2k+b +c+ 1\right) \left(2n+a +b +c +1\right) \left(2n+a+b +c +2\right)}
    S_{-}T_{+}
    \nonumber \\&
    +\frac{(k+b)(k+c)(n+k+b+c)(n+k+b+c+1)}{\left(2k+b +c+ 1\right) \left(2n+a +b +c +1\right) \left(2n+a+b +c +2\right)}
    S_{-}T_{-}. \label{HMy}
\end{align}

\subsection{Structure relations for the two-variable Jacobi polynomials}

Structure relations for continuous univariate orthogonal polynomials $p_n(x)$ are relations of the form \cite{koornwinder2007structure}:
\begin{equation}
    \pi(x) \frac{d}{dx} p_n(x) + h(x) p_n(x) =a_n p_{n+1}(x)+c_n p_{n-1}(x)
\end{equation}
where $\pi(x)$ are polynomials of degree smaller or equal to 2 and $h(x)$ is a polynomial of degree at most $1$. It should be noted that in distinction with the familiar shift relations like \eqref{uniJacobi:shift1} and \eqref{uniJacobi:shift2}, the parameters are not altered on the right hand sides of structure relations. These relations can be extended to polynomials on discrete grids with the derivative 
$\frac{d}{dx}$ replaced by the difference operator or the $q$-derivative. They were obtained in \cite{koornwinder2007structure} for the Askey-Wilson polynomials and this was shown to have a connection with the (degenerate) Sklyanin algebra in \cite{gaboriaud2020degenerate}. The extension of these characteristic relations to polynomials in two variable was examined at the general level in \cite{area2012bivariate} and \cite{fernandez2005weak}. Of direct relevance to our study is the article \cite{lewanowicz2013structure} where such structure relations are obtained for the bivariate big $q$ Jacobi polynomials and where the $q \rightarrow 1$ limit to two variable Jacobi polynomials is taken so as to produce two structure relations for these last polynomials. We shall now explain how the construction of the rank two Jacobi algebra naturally leads to these relations and more.

On the one hand we know that the non-trivial primary commutators, that is $N_1, N_3, M_3, J_1$ are first order differential operators of the form (see \eqref{Nx}, \eqref{Ny}, \eqref{My}, \eqref{Kx}, \eqref{Ky})
\begin{equation}
    f(x,y)\partial_x + g(x,y)\partial_y + h(x,y)
\end{equation}
where $f$ and $g$ are quadratic in $x$ and $y$ and $h$ is linear in the variables. On the other hand, we know the the action of these operators on the two variable Jacobi polynomials is respectively given by their realization in the degree representation as the difference operators $\Hat{N}_1, \Hat{N}_3, \Hat{M}_3, \Hat{J}_1$ whose expressions have been obtained. We thus see that equating for each case in the spirit of relation \eqref{repres} the differential action on the two-variable Jacobi polynomials with the application of the corresponding difference operators affecting the degrees, will indeed produce structure relations. In view of the fact that according to \eqref{Kx}, $J_1$  is a linear combination of $N_1, N_3, M_3$, this operator ($J_1$) will not add new relations to those obtained by using $N_1, N_3, M_3$ in the way described. Therefore only the cases corresponding to the latter set will be recorded. The findings are:

\begin{equation}
[2x(1-x) \partial_x -2xy\partial_y+ (a+1 - (a+b+c+3)x)]J_{n,k}^{(a,b,c)}(x,y)=
\Hat{N}_1 J_{n,k}^{(a,b,c)}(x,y) \label{sr1}
\end{equation}
with $\Hat{N}_1$ given by \eqref{HNx};
\begin{align}
    [2 x \left(x + y -1\right) {\partial_x}+2 y \left(x+y-1\right)\partial_y
  + \left((a + b + c+3)(x + y) -a-b-2\right)]J_{n,k}^{(a,b,c)}(x,y)= \nonumber \\
 \Hat{N}_3 J_{n,k}^{(a,b,c)}(x,y)\label{sr2}
\end{align}
with $\Hat{N}_3$ given by \eqref{HNy};
\begin{equation}
    [2y (x+y-1)\partial_y +\left((b+1)(x+y-1)+(c+1)y\right)]J_{n,k}^{(a,b,c)}(x,y)=
\Hat{M}_3 J_{n,k}^{(a,b,c)}(x,y)\label{sr3}
\end{equation}
with $\Hat{M}_3$ given by \eqref{HMy}.

\medskip

\noindent \textit{Remark}: Note that by taking the $q\rightarrow 1$ limit of the structure relations they obtained for the bivariate big $q$ Jacobi polynomials, the authors of \cite{lewanowicz2013structure} find two structure relations for the two variable Jacobi polynomials that respectively provide the actions of the differential operators $x(x+y-1)\partial_x$ and $y(x+y-1)\partial_y$ on these polynomials. Their results can be recovered as follows from those above. First note that the non-differential part on the left hand side of \eqref{sr1}, \eqref{sr2} and \eqref{sr3} can be removed with the recurrence relations \eqref{RR1} and \eqref{RR2}. Observe then that the purely differential part of \eqref{sr3} is up to a factor $y(x+y-1)\partial_y$ and that this relation thus corresponds to one of the two given in \cite{lewanowicz2013structure}; the second one is obtained as is easily seen by taking the combination: $\eqref{sr2}-\eqref{sr3}$. 
It is thus observed that the structure relations given above contain one additional independent relation, namely the action of $xy (\partial_x - \partial_y)$, which is obtained from $\eqref{sr1}+\eqref{sr2}-\eqref{sr3}$.

\section{Outlook}
In summary, we have obtained the commutation relations that result from commuting the bispectral operators of the two-variable Jacobi polynomials on the triangle. These are the differential operators $L$ and $L_1$, respectively given by \eqref{L} and \eqref{Ly} of which the polynomials $J_{n,k}^{(a,b,c)}(x,y)$ are eigenfunctions and $X_1=x$ and $X_3=1-x-y$ that are the eigenvalues of the difference operators forming the recurrence relations of these polynomials. The relations that emerged were posited to define the rank two Jacobi algebra. 
The set of generators includes the operator $L_3$ given in \eqref{Lz} which commutes with $L$.
This led to the identification of a rank one Racah algebra generated by $L_1$ and $L_3$ as a subalgebra of the rank two Jacobi algebra as well as various rank one Jacobi subalgebras; this is depicted in Figure 1.

It was further checked that, as should be, the same defining relations are recovered when the procedure just described is initiated in the degree representation where the bispectral operators are $\Hat{L}=-n\,(n+a+b+c+2)$, $\Hat{L}_1=-k\,(k+b+c+1)$ plus $\Hat{X}_1$ and $\Hat{X}_3$, the difference operators that are read out from the recurrence relations \eqref{RR1} and \eqref{RR2}.

We have recorded the expression of the various generators in this dual difference realization and noted that this provided a representation of the rank two Jacobi algebra in a basis given by the two-variable Jacobi polynomials. The combination of the differential and difference (variable and degree) realizations was seen to provide in particular structure relations for these basis vectors. Many new differential identities for the two-variable Jacobi polynomials were also provided in Appendix B.

All this suggests the examination of various fascinating questions and we shall mention a few in concluding.

A definition of the multivariate Racah algebra was already proposed through studies of the centralizer of the multifold tensor product of $sl_2$ representations \cite{de2017higher} and the representations of its rank two version have been fully constructed \cite{crampe2023representations}. It was indicated at the end of Section 3 that as for rank one, the rank two Jacobi Jacobi algebra is a specialization of the rank two Racah algebra. It would be of interest to study this connection in more depth. It is useful to remark in this respect that the Jacobi and Hahn algebras are isomorphic since the Jacobi polynomials can be obtained as limits of the Hahn polynomials and the defining relations of the algebras are oblivious to these limits. See the note on this point in Appendix A in the rank one case.

Conversely, it would be of great interest to go in the reverse direction and to determine if and how the rank two Racah algebra can be embedded in the rank two Jacobi algebra through a generalization of the tridiagonalization procedure that has been employed in \cite{genest2016tridiagonalization} to achieve this embedding in the rank one case. This should lead to the definition and characterization of two variable Wilson polynomials from the properties of the Jacobi polynomials. Pursuing in that vein as in \cite{grunbaum2017tridiagonalization} could offer a fruitful direction to study Heun-like equations in two variables.

It should be noted that the investigation presented here focused exclusively on two variable Jacobi polynomials orthogonal on the triangle. There are a number of other domains, with the disk among those, for which two variable bispectral polynomials were shown to be orthogonal \cite{koornwinder1975two}. Are there algebras that can be associated to those? 

Big $q$-Jacobi and $-1$ generalizations of the two-variable Jacobi polynomials have been given in \cite{lewanowicz2010two} and \cite{genest2015two} respectively. It should be possible to obtain, as was done here, the corresponding rank two Big $q$ and Big $-1$ Jacobi algebras that would have respectively the rank one Askey-Wilson algebra \cite{zhedanov1991hidden, crampe2021askey} and the Bannai-Ito algebra \cite{tsujimoto2012dunkl, genest2015laplace, de2015bannai} as subalgebra.

The two variable Jacobi polynomials are the wave functions of the generic superintegrable quantum model on the two sphere \cite{genest2014superintegrability,iliev2018symmetry}, this begs the question of what is the bearing of the rank two Jacobi algebra in the description of this physical model? Also, how does the study presented here relates to the classification of the rank two Leonard pairs undertaken in \cite{crampe2025factorized} with an eye in particular to superintegrable systems? Finally, one can of course envisage extending all these considerations to ranks higher than two. We are committed to examining all these questions in the near future.

\appendix

\section{Some properties of the univariate Jacobi polynomials and the rank one Jacobi algebra \label{app:A}}
We define a family of polynomials, denoted by $J_n^{(a,b)}(x)$, on the interval $[0,1]$, via the transformation $$J_n^{(a,b)}(x) = P_n^{(a,b)}(1-2x),$$ where $P_n^{(a,b)}(x)$ denotes the usual Jacobi polynomial of degree $n$, orthogonal on $[-1,1]$; see \cite{koekoek2010hypergeometric} for standard definitions and properties.
In the following, we summarize some basic properties of the polynomials $J_n^{(a,b)}(x)$.

\medskip
\noindent \textit{Explicit expression}
\begin{align*}
  &  J_n^{(a,b)}(x) = \dfrac{(a+1)_n}{n!}\hg{2}{1}\argu{-n,n+a+b+1}{a+1}{x}.
\end{align*}
\noindent \textit{Orthogonality}
\begin{align}
&   \int_{0}^1x^{a} (1-x)^{b} J_n^{(a,b)}(x) J_{m}^{(a,b)}(x)dx
     = \frac{\Gamma(n+a+1)\Gamma(n+b+1)}{(2n +a+b+1)\,n!\, \Gamma(n+a+b+1) } \delta_{nm}, \quad a,b>-1.
\end{align}

\medskip

\noindent\textit{Differential equation}
\begin{align}\label{unide}
  &H^{(a,b)}[J_{n}^{(a,b)}(x)]=-n\,(n+a+b+1)J_{n}^{(a,b)}(x),
\end{align}
where
\begin{align}
H^{(a,b)} = x(1-x)\partial_x^2 + (a+1 - (a+b+2)x)\partial_x.
\end{align}

\medskip

\noindent \textit{Recurrence relation}
\begin{align}
(1-2x) &J_n^{(a,b)}(x)
=
\frac{2(n+1)(n+a+b+1)}{(2n+a+b+1)(2n+a+b+2)}J_{n+1}^{(a,b)}(x) \nonumber  \\
&-\frac{(a-b)(a+b)}{(2n+a+b)(2n+a+b+2)}J_{n}^{(a,b)}(x)+\frac{2(n+a)(n+b)}{(2n+a+b)(2n+a+b+1)} J_{n-1}^{(a,b)}(x).
\label{RR:univariate}
\end{align}%

\medskip
\noindent \textit{Shift relation}
\begin{align}
&   \partial_x J_n^{(a,b)}(x) = -(n+a+b+1) J_{n-1}^{(a+1,b+1)}(x), \label{uniJacobi:shift1}\\
&    \left(x(1-x) \partial_x  + a-(a+b)x\right)J_n^{(a,b)}(x) = (n+1) J_{n+1}^{(a-1,b-1)}(x).
    \label{uniJacobi:shift2}
\end{align}

\medskip
\noindent \textit{Rank one Jacobi algebra} 

\noindent Take
\begin{equation}
   \ K_1=H^{(a,b)}, \qquad K_2=x,
\end{equation}
where $H^{(a,b)}$ is given by \eqref{unide} and $K_2$ is the operator multiplication by $x$. Define a third operator $K_3=[K_1,K_2]$. It is readily seen that
\begin{equation}
  K_3=2x (1-x)\partial_x +(a+1-(a+b+2)x).
\end{equation}
By direct computation one then verifies that $K_1$ and $K_2$ verify the commutation relations given in \eqref{rank1} that define the Jacobi algebra of rank one with $\alpha =a$ and $\beta =b$. We may here remark that this algebra is isomorphic to the rank one Hahn algebra (see for example \cite{frappat2019higgs}).

\medskip

\noindent \textit{Structure relation}
\begin{align}
    K_3 &J_n^{(a,b)}(x) = \dfrac{(n+1)(n+a+b+1)}{2n+a+b+1} J_{n+1}^{(a,b)}(x) -\dfrac{(n+a)(n+b)}{2n+a+b+1} J_{n-1}^{(a,b)}(x).
\end{align}

\noindent \textit{Contiguity relations of the one variable Jacobi polynomials}

\noindent We record the following contiguity relations \cite{bateman1953higher} obtained from Christoffel and Geronimus transformations \cite{chihara2011introduction} :
\begin{align}
    x J_n^{(a,b)}(x) &=\dfrac{n+a}{2n+a+b+1} J_n^{(a-1,b)}(x)-\dfrac{n+1}{2n+a+b+1} J_{n+1}^{(a-1,b)}(x),  \label{uniJacobi:CTa} \\
    J_n^{(a,b)}(x) &=\dfrac{n+a+b+1}{2n+a+b+1}J_n^{(a+1,b)}(x)-\dfrac{n+b}{2n+a+b+1}J_{n-1}^{(a+1,b)}(x), \label{uniJacobi:GTa}\\
    (1-x) J_n^{(a,b)}(x) &=\dfrac{n+1}{2n+a+b+1} J_{n+1}^{(a,b-1)}(x)+\dfrac{n+b}{2n+a+b+1} J_n^{(a,b-1)}(x),   \label{uniJacobi:CTb} \\
    J_n^{(a,b)}(x) &=\dfrac{n+a+b+1}{2n+a+b+1}J_n^{(a,b+1)}(x)+\dfrac{n+a}{2n+a+b+1}J_{n-1}^{(a,b+1)}(x). \label{uniJacobi:GTb}
\end{align}

\section{Differential properties of the two variable Jacobi polynomials \label{app:B}}
We introduce six first-order differential operators $s_1, s_2, s_3, s_1^*, s_2^*, s_3^*$ defined by
\begin{align} \label{diffop}
  &s_1=(x+y-1)\partial_y + cI, \quad
  s_2=x\partial_x + aI, \quad   
  s_3=-y\left(\partial_x-\partial_y \right)+ bI, \nonumber\\
  &s_1^*= -y \partial_y -b I, \quad
  s_2^*=(1-x-y)\partial_x-cI, \quad
  s_3^*=x(\partial_y - \partial_x) -a I,   
\end{align}
where each pair $(s_i, s_i^{*})$ for $i=1,2,3$, is related via a transformation involving parameter-dependent functions: 
\begin{align}
g_1^{(b,c)}= y^b (1-x-y)^{-c},\quad   g_2^{(a,c)}=x^{-a}(1-x-y)^c, \quad   g_3^{(a,b)}=x^a y^{-b},
\end{align}
as
\begin{align}
s_1^*=g_1^{(1-b,1-c)}\circ s_1 \circ g_1^{(b,c)}, \quad
s_2^*=g_2^{(1-a,1-c)}\circ s_2 \circ g_2^{(a,c)}, \quad
s_3^*=g_3^{(1-a,1-b)}\circ s_3 \circ g_3^{(a,b)}.    
\end{align}
The operators $L_1^{(b,c)}, L_2^{(a,c)}, L_3^{(a,b)} $ and $L$ can be expressed in terms of these six operators as follows:
\begin{align}
L_1^{(b,c)}&=(s_1^* - 1)s_1 +c(b+1)=(s_1+1) s_1^* + b(c+1), \\
L_2^{(a,c)}&=(s_2^* - 1)s_2 +a(c+1)=(s_2+1) s_2^* + c(a+1), \\
L_3^{(a,b)}&=(s_3^* - 1)s_3 +b(a+1)=(s_3+1) s_3^* + a(b+1), \label{Lzs}\\
    L&=s_1^* s_1+s_2^* s_2+s_3^* s_3 -s_1-s_2-s_3 +a+b+c+ab+ac+bc\nonumber\\
     &=s_1 s_1^*+s_2 s_2^*+s_3 s_3^* +s_1^*+s_2^*+s_3^*+a+b+c+ab+ac+bc. \label{L:s1s2s3}
\end{align}
Moreover, the actions of $s_1, s_2, s_3, s_1^*, s_2^*, s_3^*$ on the two-variable Jacobi polynomials can be readily computed and are given by:
\begin{align}
& s_1[J_{n,k}^{(a,b,c)}]=(c+k)J_{n,k}^{(a,b+1,c-1)},\label{acts2}\\
& s_1^*[J_{n,k}^{(a,b,c)}]=-(b+k)J_{n,k}^{(a,b-1,c+1)},\label{acts2*}\\
& s_2[J_{n,k}^{(a,b,c)}]=\dfrac{(n-k+1)(b+k)}{b+c+2k+1}J_{n,k-1}^{(a-1,b,c+1)}
  + \dfrac{(a+n-k)(b+c+k+1)}{b+c+2k+1}J_{n,k}^{(a-1,b,c+1)},\label{acts1}\\
& s_2^*[J_{n,k}^{(a,b,c)}]=-\dfrac{(k+1)(a+b+c+n+k+2)}{b+c+2k+1}J_{n,k+1}^{(a+1,b,c-1)}
  - \dfrac{(c+k)(b+c+k+n+1)}{b+c+2k+1}J_{n,k}^{(a+1,b,c-1)},\label{acts1*}\\
& s_3[J_{n,k}^{(a,b,c)}]= - \dfrac{(k+1)(a+b+c+n+k+2)}{b+c+2k+1}J_{n,k+1}^{(a+1,b-1,c)}
  + \dfrac{(b+k)(b+c+k+n+1)}{b+c+2k+1}J_{n,k}^{(a+1,b-1,c)},\label{acts3}\\
& s_3^*[J_{n,k}^{(a,b,c)}]=\dfrac{(n-k+1)(c+k)}{b+c+2k+1}J_{n,k-1}^{(a-1,b+1,c)}
  - \dfrac{(a+n-k)(b+c+k+1)}{b+c+2k+1}J_{n,k}^{(a-1,b+1,c)} \label{acts3*}.
\end{align}
The above relations can be proved using the contiguity relations \eqref{uniJacobi:CTa}–\eqref{uniJacobi:GTb} of the one variable Jacobi polynomials, together with additional identities: 
\begin{align}
    (x-1)\partial_x J_n^{(a,b)}(x) &= (n+a+b+1) \left(J_n^{(a+1,b)}(x)  - J_n^{(a,b)}(x)\right), \label{uniJacobi:(x-1)_diff}\\
    x \partial_x J_n^{(a,b)}(x) &= (n+a+b+1) \left(J_n^{(a,b+1)}(x)-J_n^{(a,b)}(x)\right). \label{uniJacobi:x_diff}
\end{align}
We briefly outline the proof of the third relation, namely \eqref{acts1}, as a representative case:
\begin{align}
   s_2[J_{n,k}^{(a,b,c)}]=&(x \partial_x +a I)\left(J_{n-k}^{(a,b+c+2k+1)}\bigl( x\bigr) \cdot (1-x)^k\right) \cdot J_k^{(b,c)}\left(\frac{y}{1-x}\right)    \nonumber \\
   &+J_{n-k}^{(a,b+c+2k+1)}\bigl( x\bigr) \cdot (1-x)^k \cdot x \partial_x \left(J_k^{(b,c)}\left(\frac{y}{1-x}\right) \right) \nonumber \\
   =& B_1^{(1)}(x)\, J_k^{(b,c)}\left(\frac{y}{1-x}\right)+B_2^{(1)}(x)\, J_k^{(b,c+1)}\left(\frac{y}{1-x}\right)\nonumber\\
   =& B_1^{(2)}(x)\, J_{k-1}^{(b,c+1)}\left(\frac{y}{1-x}\right)+B_2^{(2)}(x)\, J_k^{(b,c+1)}\left(\frac{y}{1-x}\right)
\end{align}
where we first rewrite the expression using \eqref{uniJacobi:x_diff}, and then express the coefficients of $J_k^{(b,c)}$ and $J_k^{(b,c+1)}$ as $B_1^{(1)}$ and $B_2^{(1)}$, respectively, to obtain the third line. The fourth line is obtained by reorganizing the terms using the contiguity relation \eqref{uniJacobi:GTb}, and expressing the coefficients of $J_{k-1}^{(b,c+1)}$ and $J_k^{(b,c+1)}$ as $B_1^{(2)}$ and $B_2^{(2)}$, respectively.
Furthermore, using the relations 
\begin{align}
     n J_n^{(a,b-1)}(x) &= (n+a+b+1)(1-x) J_{n-1}^{(a+1,b+1)}(x) - (n(1-x)+b) J_{n-1}^{(a+1,b)}(x),\\
    (n+a+1)J_n^{(a,b+1)}(x) &= (n+a+b+2)J_n^{(a+1,b+1)}(x) - (n+b+1) J_n^{(a+1,b)}(x), 
\end{align}
for $B_1^{(2)}$ and $B_2^{(2)}$, which follow directly from the four contiguity relations \eqref{uniJacobi:CTa}–\eqref{uniJacobi:GTb}, we finally obtain the right-hand side of \eqref{acts1}.

\section*{Acknowledgments}
This work has been sponsored by a Québec-Kyoto cooperation grant from the Ministère des Relations Internationales et de la Francophonie of the Quebec Government.
NC is partially supported by the international research project AAPT of the CNRS.
The research of ST is supported by JSPS KAKENHI (Grant Number 24K00528). LV is funded in part through a discovery grant of the Natural Sciences and Engineering Research Council (NSERC) of Canada. 
%The work of AZ is supported by the National Science Foundation of China (Grant No.11771015).


\begin{thebibliography}{10}

\bibitem{koekoek2010hypergeometric}
Roelof Koekoek, Peter~A Lesky, and Ren{\'e}~F Swarttouw.
\newblock {\em Hypergeometric orthogonal polynomials}.
\newblock Springer, 2010.

\bibitem{genest2016tridiagonalization}
Vincent Genest, Mourad Ismail, Luc Vinet, and Alexei Zhedanov.
\newblock {Tridiagonalization of the hypergeometric operator and the Racah--Wilson algebra}.
\newblock {\em Proceedings of the American Mathematical Society}, 144(10):4441--4454, 2016.

\bibitem{zhedanov1988nature}
AS~Zhedanov and Ya~A Granovski{\i}.
\newblock {Nature of the symmetry group of the $6j$-symbol}.
\newblock {\em Zh. Eksp. Teor. Fiz.}, 94:49, 1988.

\bibitem{genest2014superintegrability}
Vincent~X Genest, Luc Vinet, and Alexei Zhedanov.
\newblock {Superintegrability in two dimensions and the Racah--Wilson algebra}.
\newblock {\em Letters in Mathematical Physics}, 104:931--952, 2014.

\bibitem{genest2014racah}
Vincent~X Genest, Luc Vinet, and Alexei Zhedanov.
\newblock {The Racah algebra and superintegrable models}.
\newblock In {\em Journal of Physics: Conference Series}, volume 512, page 012011. IOP Publishing, 2014.

\bibitem{zhedanov1991hidden}
Alexei~S Zhedanov.
\newblock {“Hidden symmetry” of Askey-Wilson polynomials}.
\newblock {\em Theoretical and Mathematical Physics}, 89(2):1146--1157, 1991.

\bibitem{granovskii1993hidden}
Ya~I Granovskii and Alexei~S Zhedanov.
\newblock {Hidden Symmetry of the Racah and Clebsch-Gordan Problems for the Quantum Algebra $sl_q (2)$}.
\newblock {\em arXiv preprint hep-th/9304138}, 1993.

\bibitem{crampe2021askey}
Nicolas Cramp{\'e}, Luc Frappat, Julien Gaboriaud, Lo{\"\i}c~Poulain d’Andecy, Eric Ragoucy, and Luc Vinet.
\newblock {The Askey--Wilson algebra and its avatars}.
\newblock {\em Journal of Physics A: Mathematical and Theoretical}, 54(6):063001, 2021.

\bibitem{granovskii1992mutual}
Ya~I Granovskii, IM~Lutzenko, and AS~Zhedanov.
\newblock Mutual integrability, quadratic algebras, and dynamical symmetry.
\newblock {\em Annals of Physics}, 217(1):1--20, 1992.

\bibitem{lutsenko1992jacobi}
IM~Lutsenko.
\newblock {Jacobi algebra and potentials generated by it}.
\newblock {\em Theoretical and Mathematical Physics}, 93(1):1081--1090, 1992.

\bibitem{de2017higher}
Hendrik De~Bie, Vincent~X Genest, Wouter van~de Vijver, and Luc Vinet.
\newblock {A higher rank Racah algebra and the $\mathbb{Z}_{2}^{n} $ Laplace--Dunkl operator}.
\newblock {\em Journal of Physics A: Mathematical and Theoretical}, 51(2):025203, 2017.

\bibitem{crampe2021racah}
Nicolas Cramp\'e, Julien Gaboriaud, Lo{\"\i}c~Poulain d'Andecy, and Luc Vinet.
\newblock {Racah algebras, the centralizer $ Z_n(\mathfrak{sl}_2)$ and its Hilbert-Poincar\'e series}.
\newblock {\em Annales Henri Poincaré}, 23:2657--2682, 2022.

\bibitem{tratnik1991some}
MV~Tratnik.
\newblock {Some multivariable orthogonal polynomials of the Askey tableau-discrete families}.
\newblock {\em Journal of Mathematical Physics (New York)}, 32(9):2337--2342, 1991.

\bibitem{kalnins2011two}
Ernie~G Kalnins, Willard Miller, and Sarah Post.
\newblock {Two-variable Wilson polynomials and the generic superintegrable system on the 3-sphere}.
\newblock {\em SIGMA. Symmetry, Integrability and Geometry: Methods and Applications}, 7:051, 2011.

\bibitem{crampe2023representations}
Nicolas Cramp{\'e}, Luc Frappat, and Eric Ragoucy.
\newblock {Representations of the rank two Racah algebra and orthogonal multivariate polynomials}.
\newblock {\em Linear Algebra and its Applications}, 664:165--215, 2023.

\bibitem{post2024racah}
Sarah Post and S{\'e}bastien Bertrand.
\newblock {The Racah algebra of rank 2: Properties, symmetries and representation}.
\newblock {\em SIGMA. Symmetry, Integrability and Geometry: Methods and Applications}, 20:085, 2024.

\bibitem{koornwinder1975two}
Tom Koornwinder.
\newblock Two-variable analogues of the classical orthogonal polynomials.
\newblock In {\em Theory and application of special functions}, pages 435--495. Elsevier, 1975.

\bibitem{proriol1957famille}
Joseph Proriol.
\newblock Sur une famille de polyn{\^o}mes {\`a} deux variables orthogonaux dans un triangle.
\newblock {\em Comptes rendus hebdomadaires des séances de l'Académie des sciences}, 245(26):2459--2461, 1957.

\bibitem{dunkl2014orthogonal}
Charles~F Dunkl and Yuan Xu.
\newblock {\em Orthogonal polynomials of several variables}.
\newblock Number 155. Cambridge University Press, 2014.

\bibitem{karniadakis2005spectral}
George Karniadakis and Spencer~J Sherwin.
\newblock {\em {Spectral/hp element methods for computational fluid dynamics}}.
\newblock Oxford University Press, 2005.

\bibitem{krall1967orthogonal}
HL~Krall and Isador~Mitchell Sheffer.
\newblock {Orthogonal polynomials in two variables}.
\newblock {\em Annali di Matematica Pura ed Applicata}, 76:325--376, 1967.

\bibitem{engelis1974some}
GK~Engelis.
\newblock {On some two-dimensional analogues of classical orthogonal polynomials}.
\newblock {\em Latv. mat. ezhegodnik}, 15:169--202, 1974.

\bibitem{vinet2003two}
Luc Vinet and Alexei Zhedanov.
\newblock {Two-Dimensional Krall--Sheffer Polynomials and Quantum Systems on Spaces with Constant Curvature}.
\newblock {\em Letters in Mathematical Physics}, 65:83--94, 2003.

\bibitem{fernandez2010recent}
Lidia Fern{\'a}ndez, Francisco Marcell{\'a}n, Teresa~E P{\'e}rez, and Miguel~A Pinar.
\newblock Recent trends on two variable orthogonal polynomials.
\newblock {\em Differential algebra, complex analysis and orthogonal polynomials}, 5986, 2010.

\bibitem{geronimo2010bispectrality}
Jeffrey~S Geronimo and Plamen Iliev.
\newblock {Bispectrality of multivariable Racah--Wilson polynomials}.
\newblock {\em Constructive Approximation}, 31(3):417--457, 2010.

\bibitem{dunkl1980orthogonal}
Charles~F Dunkl.
\newblock {Orthogonal polynomials in two variables of $q$-Hahn and $q$-Jacobi type}.
\newblock {\em SIAM Journal on Algebraic Discrete Methods}, 1(2):137--151, 1980.

\bibitem{lewanowicz2010two}
Stanis{\l}aw Lewanowicz and Pawe{\l} Wo{\'z}ny.
\newblock {Two-variable orthogonal polynomials of big q-Jacobi type}.
\newblock {\em Journal of computational and applied mathematics}, 233(6):1554--1561, 2010.

\bibitem{lewanowicz2013structure}
Stanis{\l}aw Lewanowicz, Pawe{\l} Wo{\'z}ny, and Rafa{\l} Nowak.
\newblock {Structure relations for the bivariate big $q$-Jacobi polynomials}.
\newblock {\em Applied Mathematics and Computation}, 219(16):8790--8802, 2013.

\bibitem{genest2015two}
Vincent~X Genest, Jean-Michel Lemay, Luc Vinet, and Alexei Zhedanov.
\newblock {Two-variable $-1$ Jacobi polynomials}.
\newblock {\em Integral Transforms and Special Functions}, 26(6):411--425, 2015.

\bibitem{vinet2012limit}
Luc Vinet and Alexei Zhedanov.
\newblock {A limit $q=-1$ for the big $q$-Jacobi polynomials}.
\newblock {\em Transactions of the American Mathematical Society}, 364(10):5491--5507, 2012.

\bibitem{iliev2018symmetry}
Plamen Iliev.
\newblock Symmetry algebra for the generic superintegrable system on the sphere.
\newblock {\em Journal of High Energy Physics}, 2018(2), 2018.

\bibitem{labriet2024realisations}
Q~Labriet and L~Poulain d'Andecy.
\newblock {Realisations of Racah algebras using Jacobi operators and convolution identities}.
\newblock {\em Advances in Applied Mathematics}, 153:102620, 2024.

\bibitem{de2019bargmann}
Hendrik De~Bie, Plamen Iliev, and Luc Vinet.
\newblock {Bargmann and Barut-Girardello models for the Racah algebra}.
\newblock {\em Journal of Mathematical Physics}, 60(1), 2019.

\bibitem{GI2010}
J~S Geronimo and P~Iliev.
\newblock {Bispectrality of Multivariable Racah–Wilson Polynomials}.
\newblock {\em Constructive Approximation}, 31:417--457, 2010.

\bibitem{post2015}
Post S.
\newblock {Racah Polynomials and Recoupling Schemes of $su(1, 1)$}.
\newblock {\em Symmetry, Integrability and Geometry: Methods and Applications}, 11:057, 2015.

\bibitem{Gaboriaud2019}
J~Gaboriaud, L~Vinet, S~Vinet, and A~Zhedanov.
\newblock {The generalized Racah algebra as a commutant}.
\newblock {\em Journal of Physics: Conference Series}, 1194:012034, 2019.

\bibitem{Debie2020}
H~De Bie, P~Iliev, W~van~de Vijver, and L~Vinet.
\newblock {The Racah algebra: An overview and recent results}.
\newblock {\em Constructive Approximation}, 52:1--29, 2020.

\bibitem{LMZ2020}
D~Latini, I~Marquette, and Y-Z Zhang.
\newblock {Embedding of the Racah Algebra $R(n)$ and Superintegrability}.
\newblock {\em Annals of Physics}, 426:168397, 2020.

\bibitem{Crampe2024}
N~Crampé, L.~Frappat, J~Gaboriaud, E~Ragoucy, L~Vinet, and M~Zaimi.
\newblock {Griffiths polynomials of Racah type}.
\newblock {\em Journal of mathematical physics}, 65:083507, 2024.

\bibitem{koornwinder2007structure}
Tom~H Koornwinder.
\newblock {The structure relation for Askey--Wilson polynomials}.
\newblock {\em Journal of computational and applied mathematics}, 207(2):214--226, 2007.

\bibitem{gaboriaud2020degenerate}
Julien Gaboriaud, Satoshi Tsujimoto, Luc Vinet, and Alexei Zhedanov.
\newblock {Degenerate Sklyanin algebras, Askey--Wilson polynomials and Heun operators}.
\newblock {\em Journal of Physics A: Mathematical and Theoretical}, 53(44):445204, 2020.

\bibitem{area2012bivariate}
Iv{\'a}n Area, E~Godoy, Andr{\'e} Ronveaux, and A~Zarzo.
\newblock Bivariate second-order linear partial differential equations and orthogonal polynomial solutions.
\newblock {\em Journal of Mathematical Analysis and Applications}, 387(2):1188--1208, 2012.

\bibitem{fernandez2005weak}
Lidia Fern{\'a}ndez, Teresa~E P{\'e}rez, and Miguel~A Pi{\~n}ar.
\newblock Weak classical orthogonal polynomials in two variables.
\newblock {\em Journal of computational and applied mathematics}, 178(1-2):191--203, 2005.

\bibitem{grunbaum2017tridiagonalization}
F~Alberto Gr{\"u}nbaum, Luc Vinet, and Alexei Zhedanov.
\newblock {Tridiagonalization and the Heun equation}.
\newblock {\em Journal of Mathematical Physics}, 58(3), 2017.

\bibitem{tsujimoto2012dunkl}
Satoshi Tsujimoto, Luc Vinet, and Alexei Zhedanov.
\newblock {Dunkl shift operators and Bannai--Ito polynomials}.
\newblock {\em Advances in Mathematics}, 229(4):2123--2158, 2012.

\bibitem{genest2015laplace}
Vincent~X Genest, Luc Vinet, and Alexei Zhedanov.
\newblock {A Laplace--Dunkl equation on $S^2$ and the Bannai--Ito algebra}.
\newblock {\em Communications in Mathematical Physics}, 336:243--259, 2015.

\bibitem{de2015bannai}
Hendrik De~Bie, Vincent~X Genest, Satoshi Tsujimoto, Luc Vinet, and Alexei Zhedanov.
\newblock {The Bannai-Ito algebra and some applications}.
\newblock In {\em Journal of Physics: Conference Series}, volume 597, page 012001. IOP publishing, 2015.

\bibitem{crampe2025factorized}
Nicolas Cramp{\'e} and Meri Zaimi.
\newblock {Factorized $A_2$-Leonard pair}.
\newblock {\em The Ramanujan Journal}, 66(2):25, 2025.

\bibitem{frappat2019higgs}
Luc Frappat, Julien Gaboriaud, Luc Vinet, St{\'e}phane Vinet, and Alexei Zhedanov.
\newblock {The Higgs and Hahn algebras from a Howe duality perspective}.
\newblock {\em Physics Letters A}, 383(14):1531--1535, 2019.

\bibitem{bateman1953higher}
Harry Bateman and Arthur Erd{\'e}lyi.
\newblock {Higher transcendental functions, volume II}.
\newblock {\em Bateman Manuscript Project) Mc Graw-Hill Book Company}, 410, 1953.

\bibitem{chihara2011introduction}
Theodore~S Chihara.
\newblock {\em An introduction to orthogonal polynomials}.
\newblock Courier Corporation, 2011.

\end{thebibliography}
\end{document}